\documentclass[twocolumn,showpacs,preprintnumbers,amsmath,amssymb]{revtex4}
\usepackage[utf8]{inputenc}
\usepackage[english]{babel}
\usepackage{amsmath}
\usepackage{amsfonts}
\usepackage{amssymb}
\usepackage{graphicx}
\usepackage{lmodern}

\begin{document}

\title{Maximum power of a two-dimensional quantum mechanical engine with spherical symmetry.}
\author{J. J. Fern\'andez$^{(1)}$ and S. Omar$^{(2)}$}

\affiliation{(1) Departamento de Física Fundamental, Universidad Nacional  de Educación a Distancia (UNED), 28040 Madrid, Spain. \\
(2) Departamento de Química Física Aplicada, Universidad Autonoma de Madrid, 28049 Madrid, Spain}

\begin{abstract}
We study two-dimensional quantum Carnot engines of spherical symmetry by considering the case of a particle on the surface of a sphere of changing radius. The Carnot cycle is built allowing the state of the system to change with the specific constrains discussed in Bender's work for Carnot cycles. After studying the Carnot cycle, we maximize the output power and efficiency of the system to show that as it happens in one dimension systems:(i) the efficiency can be optimized; being its optimal value independent of the parameters describing the system and that the optimal output power at the optimal efficiency is non-zero. (ii) The optimal efficiency of the spherical system is much bigger than that of the one-dimensional quantum well considered in Abe's work.
\end{abstract} 


\date{\today}

\maketitle

\section{Introduction}

In general, thermodynamic engines convert heat into mechanical work using fluids \cite{Carnot824}. These engines absorb heat from a hot reservoir, transform a part of it into useful work and, due to their lack of perfection, reject a part of the input heat to a cold reservoir. A general definition of their efficiency is,

\begin{equation}
\eta = 1 - \dfrac{W}{Q_H}
\end{equation}

where $W$ is the mechanical work produced and $Q_H$ is the amount of heat absorbed from the hot reservoir. $Q_H$ and $W$ and $Q_c$, the heat rejected to the cold reservoir, relate by the energy conservation principle, 

\begin{equation}
Q_H = W + Q_c.
\end{equation}

As said in many works, the maximal amount of work extractable by a heat engine is gotten when the engine works reversibly and its value equals $\eta Q_H$, being $\eta$ Carnot's efficiency \cite{Carnot824}.

Carnot cycles consist in four reversible processes. In the first process the engine is in contact to a hot reservoir and undergoes through an isothermal expansion at temperature $T_H$. In the second step the machine expands adiabatically reducing its temperature to $T_C$. Then the machine goes through an isothermal compression reducing its volume while keeping its temperature equal to $T_C$. In the last step, the machine undergoes an adiabatic compression until its temperature reaches back the value $T_H$.

In several works during the last years \cite{Bender00}-\cite{Sustantyo15} some authors have explored the properties of quantum mechanical heat engines. Bender et al \cite{Bender00} considered a very simple quantum-mechanical model of a particle inside a one-dimensional square well where the movement of the walls imitates the role of a piston moving in and out. In that work the authors showed that in quantum engines it is better to use the internal energy (i.e. the expected value of the hamiltonian) instead of the temperature as the main variable to carry out the study. A main hypothesis of this study is that the efficiency of the engine, as considred to be reversible, has the form of a Carnot efficiency where the temperature is replaced by the expected (and constant) value of the Hamiltonian along the isothermal processes. Later on, S. Abe \cite{Abe11} demostrated, using the same model, that its ouput power  can be optimized and that the efficiency at maximum power has a value that does not depend on any of the parameters of the model.

In this work we extend Bender's et al and Abe's works considering a two dimensional quantum system formed by a particle of mass $m$ that is on the surface of a sphere of radius $R$ that, along the Carnot cycle lengthens and shrinks. Our work proves that Bender's formalism \cite{Bender00} can be applied succesfully to this system to obtain the expression of the efficiency $\eta$ and output power $P$. We will show that an optimal value for $\eta$ that does not depend on any parameter of the model can be found as it happens when the quantum system is the infinite well. We will also prove that the optimal output power is only a function of the radius of the sphere $R_1$ at the beggining of Carnot's cycle. 

Our work is organized as follows. In section \ref{Cycle} we present in detail the system under study and we discuss the Carnot cyle that its performs. We carry out a detailed study of the expressions of the energy, the relations among the expansion coefficients defining the system wave function and the force in each one of the steps of the Carnot cycle. Then, in section \ref{Optimizatons} we carry out an optimization of the output power of the Carnot engine. We show that the efficiency and the output power can be optimized by solving numerically a polynomial equation. In section \ref{Cremarks} we compare our results to those of Abe and present some conclusions.

\section{Theory.}
\label{Cycle}

\subsection{System description.}

Let us consider a particle of mass $m$ confined to move on the surface of a sphere of radius $R$. The time-independent Schr\"odinger equation for this system is,

\begin{equation}
\label{Hamiltonian}
-\dfrac{\hbar^2}{2m} \nabla^2 \phi(\mathbf{r}) = E \phi(\mathbf{r})
\end{equation}

where $\nabla^2 = \dfrac{1}{R^2} \mathbb{L}^2$ due to the spherical symmetry of the system and the constrain of the particle to be on the surface of a sphere of radius $R$. The eigenenergies and eigenfunctions of this Schr\"odinger equation are found in textbooks,

\begin{equation}
E_l = \dfrac{\hbar^2 l(l+1)}{2mR^2} \qquad \mathrm{and} \qquad \phi_{l}(\mathbf{r}) = \dfrac{1}{R} Y_{l0}(\theta,\phi).
\end{equation}

where $Y_{l0}(\theta,\phi)$ are complex spherical harmonics with $m=0$. Every general solution to equation (\ref{Hamiltonian}) can be written as a linear combination of such eigenfunctions

\begin{equation}
\Psi(\mathbf{r}) = \sum_l a_l \phi_{l}(\mathbf{r})
\end{equation}

where $a_l$ are complex numbers that satisfy the normalization condition $\sum_l |a_l|^2=1$.

Let us assume that the radius $R$ of the sphere can lengthen and shrink thus changing the surface of the sphere. Note that a change in $R$ leads to a change in both the eigenvalues and the eigenfunctions of the system as they are functions of $R$. As consequence, the expected value $\left< \hat{H} \right>=E(R)$ of the Hamiltonian changes with $R$. To calculate the work done along each of the processes corresponding to the Carnot cycle we need to define the force. Since in our model the parameter that changes is $R$ we define the force as $F(R)=-\dfrac{dE(R)}{dR}$.

\subsection{Carnot cycle.}

We consider a Carnot cycle where the starting point is that of a particle in the state with $l=1$ of a sphere of radius $R_1$. At this point the energy and wavefunction of the system are, 

\begin{equation}
E(R_1) = \dfrac{\hbar^2}{mR_1^2} \quad \phi_{l}(R_1) = \dfrac{1}{R} Y_{10}(\theta,\phi),  
\end{equation}

and the force acting on the system is,

\begin{equation}
F(R_1) = -\dfrac{2\hbar^2}{mR_1^3}.
\end{equation}

\begin{figure}
\includegraphics[scale=0.30]{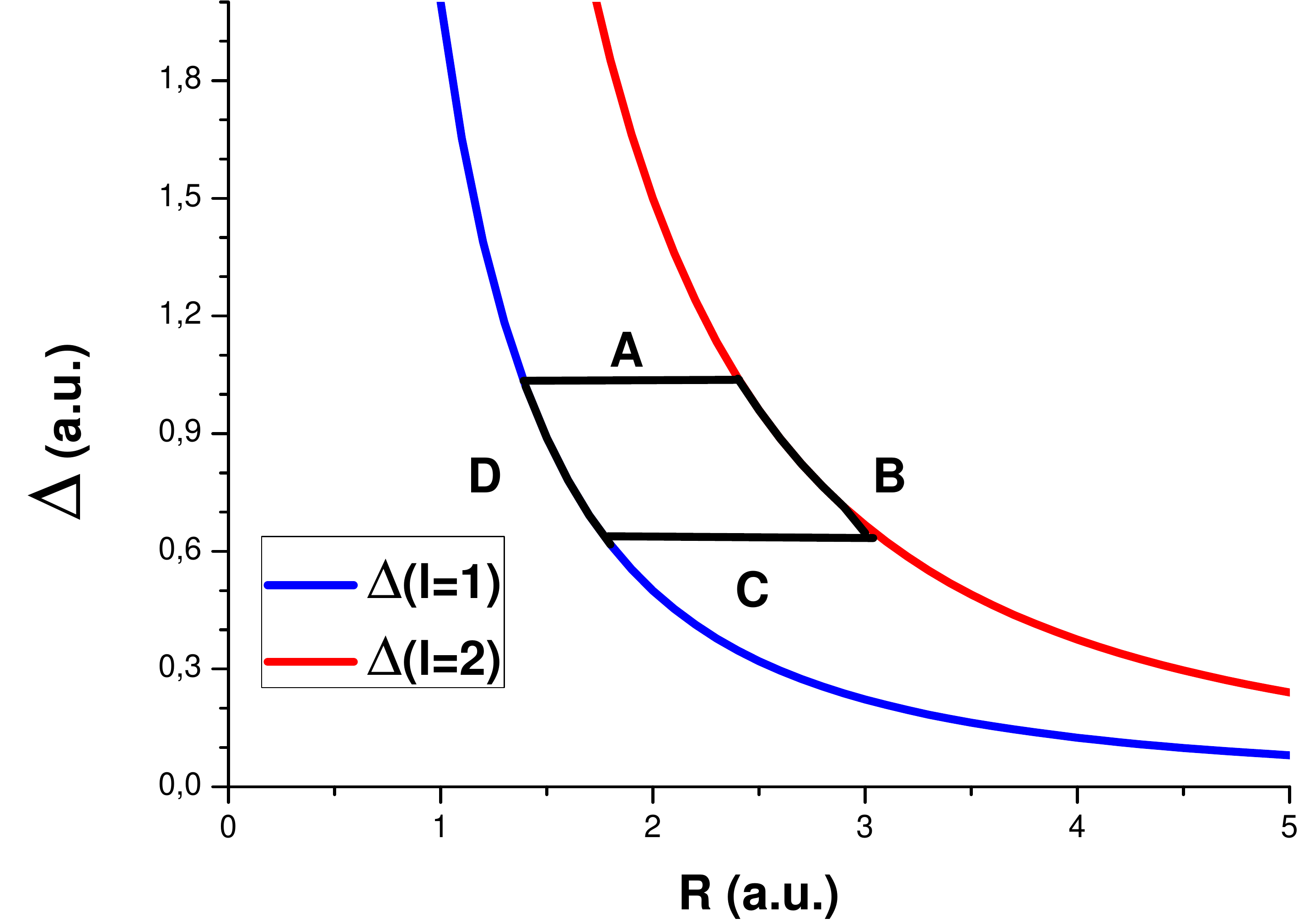}
\caption{Schematic representation of a Carnot cycle based on the expansion and compression of a system of a particle on the surface of a sphere. We detail (A) the isothermal expansion, (B) the adiabatic expansion, (C) the isothermal compression and (D) the adiabatic compression. The vertical axis is $\Delta=\dfrac{2mE_l}{\hbar^2}$ (in a.u.).}
\label{FigureCycle}
\end{figure}

In Figure \ref{FigureCycle} we show a representation of the Carnot cycle described in this section. Note that the representation is done with a starting value of the radius $R_1=1.39$ a.u 
for drawing needs.

\textit{A. Isothermal expansion.} We start Carnot's cycle by expanding the sphere isothermally. As the radius of the sphere increases we excite the system from the initial state with $l=1$ to the state with $l=2$.  A general form for the wave function for every value of $R$ along this expansion is,

\begin{equation}
\psi^A(\mathbf{r}) = a_1(R) \phi_1(\mathbf{r})|_{|\mathbf{r}|=R} + a_2(R) \phi_2(\mathbf{r})|_{|\mathbf{r}|=R}.
\end{equation}

The coefficients $a_1(R)$ and $a_2(R)$ satisfy $|a_1(R)|^2 + |a_2(R)|^2 = 1$ due to the normalization condition of the wave function for every value of $R$. The expected value of the Hamiltonian evaluated over $\psi^A(\mathbf{r})$ is,

\begin{equation}
\label{HvalueIsothermal1}
E=\left< \psi^A(\mathbf{r}) | \hat{H} | \psi^A(\mathbf{r}) \right> = |a_1|^2 \dfrac{\hbar^2}{mR^2} + |a_2|^2 \dfrac{3\hbar^2}{mR^2} = E_1(R_1)
\end{equation}

Note that the isothermal character of the expansion imposes a restriction on the expected value of the hamiltonian that must be, for every value of $R$, equal to the initial energy $E_1(R_1)$. Equation (\ref{HvalueIsothermal1}) and the normalization condition of the wave function allow to calculate $|a_1(R)|^2$ and $|a_2(R)|^2$ in terms of $R$ and $R_1$,

\begin{equation}
\label{CoefsIsothermal1}
|a_1(R)|^2 = \dfrac{1}{2} \left[ 3 - \dfrac{R^2}{R_1^2}\right] \qquad |a_2(R)|^2 = \dfrac{1}{2} \left[ \dfrac{R^2}{R_1^2} -1 \right]
\end{equation} 

Note that $|a_2(R_1)|^2=0$ satisfies the initial condition of our problem. Moreover, imposing the condition $|a_1(R_{max})|^2=0$ \cite{Abe11} we obtain that the maximum radius $R_{max}$  to which the sphere can be expanded isothermally. This value is $R_{max}=\sqrt{3} R_1$. At $R_{max}$  the level with $l=2$ is fully populated and the level with $l=1$ has an occupation equal to zero.

Using the definition of the force given at the begining of this section, $F(R)=-\dfrac{dE(R)}{dR}$, we obtain the force for each value of $R$,

\begin{eqnarray}
\nonumber
& F(R) = |a_1(R)|^2 \dfrac{d E_1}{d R} + |a_2(R)|^2 \dfrac{d E_2}{d R} = & \\
\label{ForceIsothermal1}
& |a_1(R)|^2 \dfrac{2\hbar^2}{mR^3} + |a_2(R)|^2 \dfrac{6\hbar^2}{mR^3} &   
\end{eqnarray}

Finally, we calculate the work done on the system by integrating the force, eq. (\ref{ForceIsothermal1}), from $R=R_1$ to $R=\sqrt{3}R_1$. The result of this integral is 

\begin{equation}
W^A=\dfrac{\hbar^2}{mR_1^2} \mathrm{Ln}3.
\end{equation}

\textit{B. Adiabatic expansion.} The second step of Carnot's cycle involves an adiabatic expansion from $\sqrt{3}R_1$ to $R_3$. Note that at this point of the calculation $R_3$ is still unknown and that its value will be given later. Since the expansion is adiabatic the system state does not change so, in every moment it is,
$\psi^B(\mathbf{r})= \phi_2(\mathbf{r})$ and the energy of the system is given by,

\begin{equation}
E^B(R) = \dfrac{3\hbar^2}{mR^2}
\end{equation}

Note that $R$ changes along the expansion so it does $E^B(R)$. As in step A, we calculate the force as $F(R)=-\dfrac{dE(R)}{dR}$ and the work as the integral of the force over $R$. The limits for $R$ are $R_{min}=\sqrt{3}R_1$ and $R_{max}=R_3$. The result for the work is,

\begin{equation}
W^B = \dfrac{\hbar^2}{m R_1^2} - \dfrac{6\hbar^2}{m R_3^2}.
\end{equation}

\textit{C. Isothermal compression.} 

The third step in the Carnot's cycle is an isothermal compression where the system radius decreases from $R_3$ to the value $R_{min}$ to be determined. In this process the system energy keeps constant and equal to the value $E=E^B(R_3)=\dfrac{3\hbar^2}{mR_3^2}$. The isothermal compression starts with the system in state $\psi^C(\mathbf{r})= \phi_2(\mathbf{r})$ with $|r|=R_3$ and along the compression it takes the general form,

\begin{equation}
\psi^C(\mathbf{r}) = b_1(R) \dfrac{Y_{10}(\theta,\phi)}{R} + b_2(R) \dfrac{Y_{20}(\theta,\phi)}{R}
\end{equation}

As in step $A$, the normalization of the wave function leads to the condition $|b_1(R)|^2+|b_2(R)|^2=1$. Moreover, since the compression is carried out isothermally, we have the following restriction on the expected value of $\hat{H}$,

\begin{equation}
\label{Isothermal2}
E_2(R_3)=\dfrac{3\hbar^2}{mR_3^2} = |b_1(R)|^2 E_1(R) +|b_2(R)|^2 E_2(R)
\end{equation} 

Using the normalization condition and the restriction on the energy, eq. (\ref{Isothermal2}), we obtain the coefficients $|b_1(R)|^2$ and $|b_2(R)|^2$,

\begin{equation}
|b_1(R)|^2=\dfrac{3}{2} \left[ 1 - \dfrac{R^2}{R_3^2} \right] \qquad |b_1(R)|^2=\dfrac{1}{2} \left[ \dfrac{3R^2}{R_3^2}-1 \right]
\end{equation}

Note that $|b_1(R_3)|^2=0$ and $|b_2(R_3)|^2=1$ satisfy the initial condition of the wave function since the compression starts with the system in the state $\phi_2(\mathbf{r})$ and with $|\mathbf{r}|=R_3$. To calculate the minimum radius that can be reached in this compression we impose the condition $|b_2(R_{min})|^2=0$. We obtain the value $R_{min} = R_3/\sqrt{3}$. At $R_{min}$ the wave function
of the system is $\phi_1(\mathbf{r})$ (with $|\mathbf{r}|=R_3/\sqrt{3}$) and the energy is $E=\dfrac{3\hbar^2}{mR_{3}^2}$.

As in the other steps of the Carnot cycle we  calculate the force for every $R$ using $F(R)=-|b_1|^2 \dfrac{d E_1(R)}{d R} -|b_2|^2 \dfrac{d E_2(R)}{d R}$. The result is,

\begin{equation}
F^C(R) = \dfrac{6 \hbar^2}{m R_3^2 R}.
\end{equation}

Integrating $F^C(R)$ from $R_3$ to $R_3/\sqrt{3}$, we obtain the work corresponding to this step of the Carnot cycle

\begin{equation}
W^C = \dfrac{3\hbar^2}{mR_3^2} \mathrm{Ln}3.
\end{equation}

\textit{D. Adiabatic compression.} To end the Carnot cycle we perform an adiabiatic compression where the radius of the sphere shrinks from $R_3/\sqrt{3}$ to $R_1$. Due to the adiabaticity of the process, the system remains for every value of $R$ in the state with $l=1$. The expected value of the Hamiltonian is $E^D=\dfrac{\hbar^2}{mR^2}$. From $E^D(R)$ we obtain the value of the force,

\begin{equation}
F^D(R) = \dfrac{2\hbar^2}{mR^3}.
\end{equation}

Once the system ends process \textit{D} it comes back to its initial state where it is described by the wave function $\phi_1(\mathbf{r})$ (with $|\mathbf{r}|=R_1$) and it has the initial energy $E_1(R_1) = \dfrac{\hbar^2}{mR_1^2}$.

As in every Carnot cycle, in the two isothermal processes the system exchanges heat with its surroundings. Since in those two processes the expected value of the Hamiltonian remains constant (and equal to $E^A$ and $E^C$ respectively) we know that the work received or done by the system equals the amount of heat that it receives or gives to its surroundings. That is, we know that the engine absorbs the heat 

\begin{equation}
Q_H = -W^A = -\dfrac{\hbar^2}{mR_1^2} \mathrm{Ln}3
\end{equation}

during the isothermal expansion and rejects the heat

\begin{equation}
Q_c = W^C = \dfrac{\hbar^2}{mR_3^2} \mathrm{Ln}3
\end{equation}

during the isothermal compression. The total amount of work done by the system during the whole Carnot cycle is given by,

\begin{equation}
W = -W^A-W^B+W^C+W^D = \dfrac{\hbar^2}{m} \left[ \dfrac{ \mathrm{Ln}3+2}{R_1^2} - \dfrac{9-3\mathrm{Ln}3}{R_3^2} \right]
\end{equation}

In the next section we find out the value of $R_3$.

\section{Efficiency optimization.}
\label{Optimizatons}

We now look for the value of $R_3$ and the best efficiency $\eta^*$ that can be obtained with the Carnot cycle described in section \ref{Cycle}.B. The efficiency of the Carnot cycle is defined as the ratio of the work done $W$ to the heat $Q_H$ absorbed  

\begin{equation}
\eta = \dfrac{W}{Q_H}=1-\dfrac{E_2(R_3)}{E_1(R_1)} = 1 -  \dfrac{3R_1^2}{R_3^2} = 1 - \dfrac{3}{\lambda^2}
\end{equation}

where in the last equality we have introduced the parameter $\lambda=R_3/R_1$ as is done in Abe's work \cite{Abe11}. Let us note that $\eta \geq 0$ so $\lambda \geq \sqrt{3}$, or in terms of the spheres radii $R_3 \geq \sqrt{3} R_1$.

As explained in Abe's work, it is possible to find out a value of $\lambda$ that optimizes the Carnot cycle performed by the system optimizing the output power. We start assuming that the change of the radius $\Delta R$ of the sphere happens at constant speed  $v$. This condition is equal to admit that  $\Delta R = v \cdot t_{tot}$ being the value of $\Delta R=2(R_3-R_1)=2R_1(\lambda-1)$ and $t_{tot}$ the total time used to perform the cycle \cite{Abe11}. Having $t_{tot}$ we define the output power as,

\begin{equation}
P(\lambda) = \dfrac{W(\lambda)}{t_{tot}} = \dfrac{\hbar^2 v}{2mR_1^3} \dfrac{(2+\mathrm{Ln}3) \lambda^2-3(3-\mathrm{Ln}3)}{\lambda^2(\lambda-1)}
\end{equation} 

Note that we have expressed $W$, and thus $P$, as function of $\lambda$ since this is the relevant parameter for the optimization. The value of $\lambda$ optimizing $P(\lambda)$ is obtained by solving the equation $d P(\lambda)/d \lambda =0$ or,

\begin{equation}
\label{Maximum-r}
\dfrac{-(\mathrm{Ln}3+2)\lambda^3-9r\mathrm{Ln}3+27\lambda-6\mathrm{Ln}3-18}{\lambda^3(\lambda-1)} = 0.
\end{equation}

Equation (\ref{Maximum-r}) has three real solutions  $\lambda_1=-3.74484$, $\lambda_2=0.69485$ and $\lambda_3=3.74484$. Of them only $\lambda_3$ has physical significance since $\lambda \geq \sqrt{3}$. Note that for this value of $\lambda$, $R_3$ takes the value $R_3^*=3.74484 R_1$, the efficiency is $\eta^*=0.78607$ and the output power is, $P^*=1.043\dfrac{\hbar^2 v}{2mR_1^3}$.

\section{Final remarks}
\label{Cremarks}

In this work we have studied a 
two-dimensional Quantum Carnot engine with 
spherical symmetry and we have optimized its 
efficiency and output power. We have shown that it 
is possible to define exactly the state of the 
system and its energy at every point of each one 
of the steps of the Carnot cycle. Using conditions 
on the efficiency we have then shown that every 
Carnot cycle done by the sphere must involve state 
that, after the two expansions, has a radius that is at least $
\sqrt{3}$ times larger than the initial radius. 
Otherwise the Carnot cycle is wrongly defined and 
its efficiency turns negative. 

In the last part of the work we have 
defined and optimized the output power of our two-dimensional Carnot engine. The optimization gives that the optimal efficiency (0.78607) and output powers ($P^*=0.521\dfrac{\hbar^2 v}{mR_1^3}
$, in atomic units) are obtained for a cycle where 
$\lambda=R_3/R_1=3.74484$. 

To end this work, we compare our results to those obtained in previous works. In Abe's work \cite{Abe11}, we find that in a one-dimension infinite well the optimal value of $\lambda$ (that is defined as the ratio between $L_3$, the maximum length that has the well after de isothermal and adiabatic expansions and the initial length $L_1$) is 3.0641. We see that in our work the parameter $\lambda$ is slightly larger since $\lambda=3.74484$. Abe also finds that the optimal output power is $P^*(Abe)=0.9511 \dfrac{\hbar^2 v}{mL_1^3}$, we find that in the spherical system $P^*=0.521\dfrac{\hbar^2 v}{mR_1^3}
$. If we set $R_1 = L_1$ (note that both amounts are a scale parameter for the output power), we see that the output power obtained in our work is about the half of the output power obtained in Abe's work. Finally, we see that the optimal efficiency obtained in this work (0.7861) is much larger than that obtained for the one-dimensional infinite well (0.4260). Note that the results discussed here prove that in quantum Carnot cycles the optimal efficiency depends closely on the form of the eigenvalues of the Hamiltonian describing the problem. That is, the efficiency and output power of the cycle are functions depending on the dimensionality and the symmetry of the problem.

\section*{Acknowledgements.}

The authors thank to the UNED for for providing its computational facilities to carry out the work and to Prof. J. E. Alvarellos and Drs. David Garc\'ia Aldea, Eva M. Fern\'andez and J. Rodr\'iguez-Laguna for fruitful discussions.

\end{document}